\documentclass[aps,twocolumn,amsmath,letterpaper,floatfix]{revtex4}
\usepackage{amssymb}
\usepackage{graphicx}
\usepackage{array}
\usepackage{hhline}
\usepackage{longtable}

\newcommand\tp{\rule{0pt}{2.6ex}}
\newcommand\bt{\rule[-1.2ex]{0pt}{0pt}}

\begin{document}

\title{Tensor Renormalization Group: Local Magnetizations, Correlation Functions, and

Phase Diagrams of Systems with Quenched Randomness}

\author{Can G\"uven$^{1,2}$, Michael Hinczewski$^{3,4}$, and A. Nihat Berker$^{5,6}$}
\affiliation{$^1$Department of Physics, University of Maryland,
College Park, Maryland 20742, U.S.A.,} \affiliation{$^2$Department
of Physics, Ko\c{c} University, Sar\i yer, Istanbul 34450, Turkey,}
\affiliation{$^3$Feza G\"ursey Research Institute, T\"UBITAK -
Bosphorus University, \c{C}engelk\"oy, Istanbul 34684, Turkey,}
\affiliation{$^4$Institute for Physical Science and Technology,
University of Maryland, College Park, Maryland 20742, U.S.A.,}
\affiliation{$^5$Sabanc{\i} University, Faculty of Engineering and
Natural Sciences, Orhanl\i -Tuzla, Istanbul 34956, Turkey,}
\affiliation{$^6$Department of Physics, Massachusetts Institute of
Technology, Cambridge, Massachusetts 02139, U.S.A.}

\begin{abstract}
The tensor renormalization-group method, developed by Levin and
Nave, brings systematic improvability to the position-space
renormalization-group method and yields essentially exact results
for phase diagrams and entire thermodynamic functions.  The method,
previously used on systems with no quenched randomness, is extended
in this study to systems with quenched randomness.  Local
magnetizations and correlation functions as a function of spin
separation are calculated as tensor products subject to
renormalization-group transformation.  Phase diagrams are extracted
from the long-distance behavior of the correlation functions. The
approach is illustrated with the quenched bond-diluted Ising model
on the triangular lattice.  An accurate phase diagram is obtained in
temperature and bond-dilution probability, for the entire
temperature range down to the percolation threshold at zero
temperature.

\end{abstract}

\maketitle
\def\s{\rule{0in}{0.28in}}
\setlength{\LTcapwidth}{\columnwidth}

\begin{figure}[t]
\includegraphics*[width=\columnwidth]{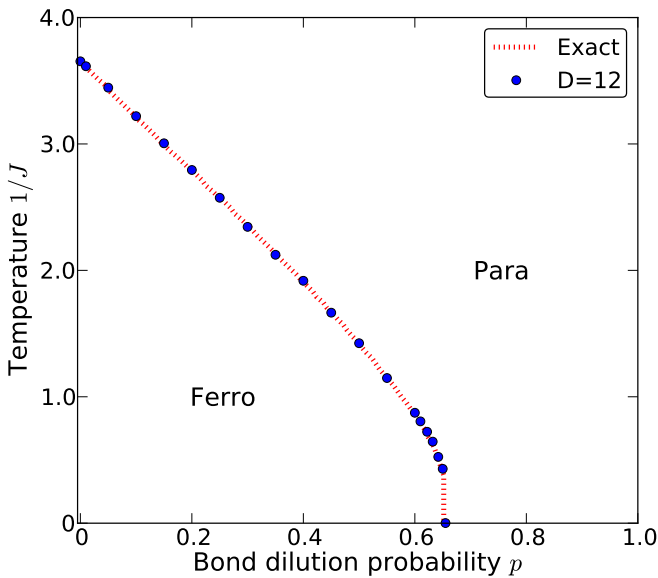}
\caption{\label{fig6} (Color online) The phase diagram of the
bond-diluted Ising model on a triangular lattice, showing the
transition temperature as a function of the bond dilution
probability $p$.  The ferromagnetic (Ferro) and paramagnetic (Para)
phases are marked. The phase boundary line between these two phases
connects, at zero temperature, with the percolation transition on
the triangular lattice.  Filled circles are our results using the
TRG method with $D=12$ together with finite size scaling, as
described in Sec.~\ref{results}.  The red dotted line is the result
of Georges {\it et al.}~\cite{Georges}, which is exact on the scale
of the figure.}
\end{figure}

\section{Introduction}

The tensor renormalization-group (TRG) method developed by Levin and
Nave~\cite{LN} is a highly useful update of the traditional
position-space renormalization-group approaches. While these
founding approaches relied on uncontrolled approximations that were
often
system-specific~\cite{NvL,Migdal,Kadanoff,Kadanoff1,Kadanoff2,Berker},
the TRG is general in scope---it works on any classical
two-dimensional lattice Hamiltonian with local interactions---and
its accuracy can be systematically improved to converge on the exact
thermodynamic results.  Along with these advantages, the method fits
within the conceptual framework of traditional renormalization-group
theory: it is a mapping between Hamiltonians on the original and
coarse-grained lattices, and phase transition behavior can be
extracted from flows of the Hamiltonians as the transformation is
iterated~\cite{HinczBerkerTRG}.

The initial TRG study demonstrated the power of the approach in the
context of the triangular-lattice Ising model~\cite{LN}. Since then
it has proven a versatile tool for a variety of classical systems,
including the frustrated Ising model on a Shastry-Sutherland
lattice~\cite{Chang}, relevant to magnetization plateaus in
rare-earth tetraborides, and the zero-hopping limit of a model for
ultra-cold bosonic polar molecules on a hexagonal optical
lattice~\cite{Bonnes}. Moreover, the ideas behind the TRG method
have become the kernel for developments in two-dimensional quantum
systems~\cite{Gu,Jiang,Xie,Zhao,Chen,Li}, most notably
tensor-entanglement renormalization group for studying symmetry
breaking and topological phase transitions~\cite{Gu}, and accurate
methods to calculate ground-state expectation
values~\cite{Jiang,Xie,Zhao}. Beyond the precision of the method, a
key factor spurring the growth of tensor RG applications in both
classical and quantum cases is computational efficiency: the CPU
cost of carrying out TRG scales linearly with lattice
size~\cite{Zhao}.

Given these promising characteristics, TRG is a natural candidate
for tackling models with quenched randomness---a field where
extracting accurate phase diagram information is a significant
challenge.  The current study presents the first example of TRG
applied to such a system with frozen disorder, namely the
percolative system of the bond-diluted triangular-lattice Ising
ferromagnet, yielding, as seen in Fig.1, a highly accurate global
phase diagram, down to zero temperature, where it connects with the
percolation transition.

Our paper is organized as follows: Sec.~\ref{method} develops the
TRG method for a general quenched random system. Sec.~\ref{Ising}
illustrates this tensor network mapping in particular for the
bond-diluted model and shows how to extract physical observables
such as spin-spin correlation functions. Sec.~\ref{results} uses
this method, together with finite-size scaling relations for the
correlation functions, to derive our main result: The phase diagram
in terms of temperature vs. bond dilution probability.  Close
agreement with the known critical temperature curve~\cite{Georges}
is achieved even at a relatively low order of the TRG approximation
(i.e., a small cutoff parameter).  Our work opens up future
possibilities for the extensive use of TRG in quenched disordered
systems, as argued in the concluding remarks of
Sec.~\ref{conclusions}.

\section{TRG method for quenched random systems}\label{method}

\subsection{The Tensor Network}

As in earlier studies~\cite{LN,HinczBerkerTRG}, we focus here on
classical Hamiltonians associated with hexagonal-lattice tensor
networks, though the method that we develop for quenched random
systems is readily generalized to other geometries like the square
and kagom\'e lattices~\cite{LN}. We consider a general Hamiltonian
that involves local interactions expressed in terms of bond degrees
of freedom, such that each bond has $d$ possible states and the
partition function of the system has the form
\begin{equation}\label{m1}
Z = \sum_{i_1,\ldots,i_{K}=1}^d T_{i_1 i_2 i_3} T_{i_3 i_4 i_5}
\cdots T_{i_{K-2} i_{K-1} i_{K}},
\end{equation}
where, for each of the $N$ sites in the hexagonal lattice, a
real-valued tensor $T_{i_m i_n i_o}$ is a Boltzmann weight depending
on the configuration of the three bonds meeting at the site.  The
bond degrees of freedom correspond to each tensor index running from
1 to $d$.  These bond indices are labeled $i_1$ through $i_{K}$, for
the total of $K = 3N/2$ bonds in the lattice. Although the tensor
can have as many as $d^3$ distinct non-zero elements, in practice
some bond configurations may be disallowed for a given Hamiltonian,
corresponding to zero-valued tensor elements.

\begin{figure}
\includegraphics*[width=\columnwidth]{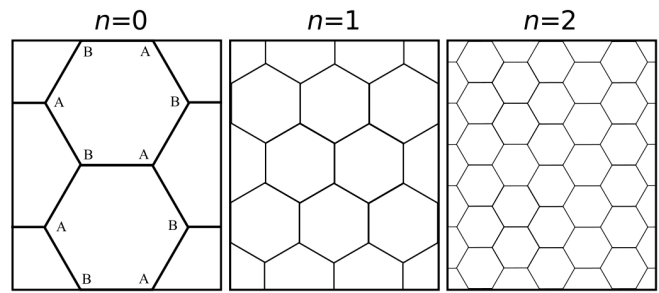}
\caption{\label{fig1} Construction of the hexagonal lattice used in
the TRG procedure. Starting from the initial structure on the left
($n=0$), at each construction step, every vertex is replaced by a
hexagon. Periodic boundary conditions are imposed between the top
and bottom edges and between the left and right edges, as if the
lattice is on the surface of a torus.  The sublattices $A$ and $B$
are shown for the $n=0$ step.}
\end{figure}

\begin{figure*}
\includegraphics*[width=\textwidth]{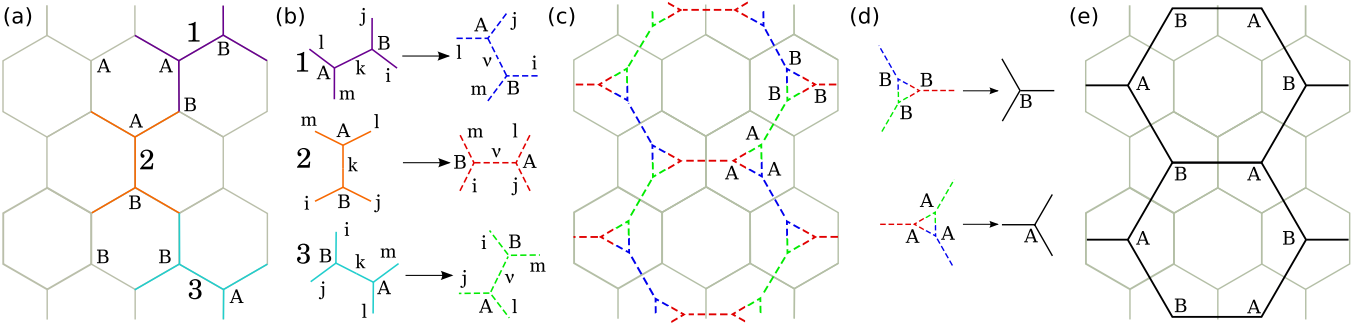}
\caption{\label{fig2} (Color) The TRG transformation described in
Sec.~\ref{TRGtrans}.  (a) The hexagonal tensor network, with the
three representative orientations of $T^A$ and $T^B$ tensor pairs,
labeled as cases 1 through 3 and highlighted in different colors.
(b) For each of the three cases, the rewiring step [Eq.~\eqref{t1}],
expressing the contraction equivalently in terms of different
tensors $S^A$ and $S^B$.  (c) After every pair of tensors is
rewired, the resulting martini lattice of $S^A$ and $S^B$ tensors.
The original lattice is superimposed in gray for reference. (d) The
decimation step [Eq.~\eqref{t2}], which replaces three $S^A$ tensors
by a renormalized $T^{\prime A}$ tensor (and analogously for $S^B$).
(e) The final lattice of renormalized $T^{\prime A}$ and $T^{\prime
B}$ tensors, with the original lattice in gray for comparison.}
\end{figure*}

To facilitate the description of the TRG procedure, the hexagonal
lattice is constructed as illustrated in Fig.~\ref{fig1}: At the
$n$th step, we replace each vertex with a hexagon, with the initial
structure denoted $n=0$.  We impose periodic boundary conditions,
such that the top and bottom edges are equivalent, as well as the
left and right edges, so that the lattice effectively lies on the
surface of a torus.  After the $n$th step, the system has $N=8\cdot
3^n$ sites and $K= 4\cdot 3^{n+1}$ bonds. The TRG method involves a
renormalization-group transformation that reverses this construction
process, mapping the system at step $n$ to one at step $n-1$.

The hexagonal lattice of any size can be decomposed into two
sublattices $A$ and $B$, such that the nearest neighbors of one type
belong to the other type.  As an example, we label the sublattices
in the $n=0$ panel of Fig.~\ref{fig1}.  We distinguish the
sublattice tensors with superscripts, $T^A_{i_m i_n
  i_o}$ or $T^B_{i_m i_n i_o}$.  In the
partition function sum of Eq.~\eqref{m1}, each bond index $i_m$
appears twice, once within an $A$ tensor, and once within the
neighboring $B$ tensor linked through that bond.  Thus evaluating $Z$
consists of performing $K$ tensor contractions.

In addition to the bond variables, the general system we consider
has quenched random degrees of freedom, though for notational
simplicity we shall not explicitly show the dependence of $T$ on
these.  Physical observables $Q$ will be expressed as $[ \langle Q
  \rangle ]$, where $\langle \cdot \rangle$ denotes the thermodynamic
average over the bond degrees of freedom and $[ \cdot ]$ denotes the
configurational average over the quenched disorder.

\subsection{The TRG Transformation}\label{TRGtrans}

The TRG transformation consists of two steps, known as rewiring and
decimation.  In the rewiring step, the bonds of every pair of
neighboring tensors $T^A$ and $T^B$ are reconnected, rewriting them
as a contraction of two new tensors $S^A$ and $S^B$.  The
reconnection pattern is illustrated in Fig.~\ref{fig2}(a,b) and can
be broken down into three basic cases (highlighted in different
colors) involving different orientations of the intial $T^A$ and
$T^B$ tensors.  In our graphical convention, the vertex where three
solid lines meet is a $T$ tensor and the vertex where three dashed
lines meet is an $S$ tensor. Indices on a tensor, i.e., $T^A_{ijk}$,
correspond to bonds labeled $i$, $j$, $k$ arranged counterclockwise
around the tensor, with the first index marking the vertical bond
for the $T$ tensors and the horizontal bond for the $S$ tensors.
Thus for example the three rewirings shown in Fig.~\ref{fig2}(b)
denote the mathematical identities
\begin{equation}\label{t1}
\begin{split}
\text{Case 1}:\qquad \sum_{k=1}^d T^A_{mkl}T^B_{jki} &=
\sum_{\nu=1}^{d^2}
S^A_{l\nu j}S^B_{i\nu m},\\
\text{Case 2}:\qquad \sum_{k=1}^d T^A_{klm}T^B_{kij} &=
\sum_{\nu=1}^{d^2} S^A_{\nu j l}S^B_{\nu m i},\\
\text{Case 3}:\qquad \sum_{k=1}^d T^A_{lmk}T^B_{ijk} &=
\sum_{\nu=1}^{d^2} S^A_{jl\nu}S^B_{m i\nu}.
\end{split}
\end{equation}
Note that the $S$ tensors have two indices which run up to $d$
(labeled by Latin letters) and one index that runs up to $d^2$
(labeled by a Greek letter).  The reason why $S^A$ and $S^B$ must
have this structure comes from the following derivation, which also
illustrates how one can explicitly calculate these tensors.

We shall describe the derivation for case 1, since the other two
cases are analogous.  The first line of Eq.~\eqref{t1} can be
expressed as a $d^2 \times d^2$ matrix equation, $M = S^A (S^B)^T$,
where $M_{\alpha\beta} \equiv \sum_k T^A_{mkl}T^B_{jkl}$,
$S^A_{\alpha\nu} \equiv S^A_{l\nu j}$, $S^B_{\beta\nu} \equiv
S^B_{i\nu m}$.  Here we use composite indices $\alpha$ and $\beta$
with $d^2$ states defined as $\alpha \equiv (j,l)$ and $\beta \equiv
(m,i)$.  As a real-valued matrix, $M$ has a singular value
decomposition of the form $M = U \Sigma V^T$, where $U$, $V$ are
orthogonal matrices and $\Sigma$ is a diagonal matrix containing the
$d^2$ singular values of $M$.  Once the singular value decomposition
of $M$ is calculated, the elements of $S^A$ and $S^B$ are given by
$S^A_{\alpha\nu} = \sqrt{\Sigma_{\nu\nu}} U_{\alpha\nu}$,
$S^B_{\beta\nu} = \sqrt{\Sigma_{\nu\nu}} V_{\beta\nu}$, where
$\Sigma_{\nu\nu}$ is the $\nu$th singular value, adopting the
ordering convention from largest to smallest with increasing $\nu$.

After all $T^A$ and $T^B$ pairs are rewired, we have a so-called
martini lattice of $S^A$ and $S^B$ tensors, shown in
Fig.~\ref{fig2}(c).  The final step of the TRG transformation is
decimation, which traces over the degrees of freedom in the
triangles of the martini lattice, substituting for each triangle a
renormalized tensor $T^{\prime A}$ or $T^{\prime B}$.  Graphically,
Fig.~\ref{fig2}(d) shows the decimation of three $S^A$ tensors to
form $T^{\prime A}$ and of three $S^B$ tensors to form $T^{\prime
B}$. The corresponding expressions in terms of tensor components are
\begin{equation}\label{t2}
\begin{split}
\sum_{j,l,h=1}^d S^A_{\nu j l}S^A_{l\gamma h}S^A_{h j \delta} &=
T^{\prime A}_{\nu\gamma\delta},\\
\sum_{m,i,h=1}^d  S^B_{\nu m i} S^B_{i\gamma h} S^B_{h m\delta} &=
T^{\prime B}_{\nu\gamma\delta}.
\end{split}
\end{equation}
The final renormalized tensor network of $T^{\prime A}$ and $T^{\prime
  B}$ is shown in Fig.~\ref{fig2}(e).

The partition function $Z$, a contraction over all bonds connecting
the tensors, Eq.~\eqref{m1}, is exactly preserved through this
transformation, as the hexagonal lattice is coarse-grained from a
step $n$ to a step $n-1$ structure.  However, the indices of the
renormalized tensors run from 1 to $d^2$ instead of 1 to $d$, so
that if the TRG were iterated, arbitrarily large tensors would
result, making numerical implementation difficult.  This problem is
related to a general feature of position-space renormalization on
lattices: except for specially tailored geometries (i.e.,
hierarchical lattices~\cite{BerkerOstlund,Kaufman,Kaufman2}), the
number of couplings in the renormalized Hamiltonian grows with each
coarse-graining.  For the TRG, we can tackle this issue in a
systematic fashion by truncating the index range with an upper bound
$D$.  In Eq.~\eqref{t2} for $T^{\prime A}$ and $T^{\prime B}$, we
shall allow the indices $\nu$,$\gamma$, and $\delta$ to run only up
to $\bar{d} \equiv \min(d^2,D)$.  This is equivalent to using
truncated matrices $\bar{S}^A$ and $\bar{S}^B$ in the rewiring step,
where $\bar{S}^A$ is the first $\bar{d}$ columns of the $d^2 \times
d^2$ matrix $S^A$ and $\bar{S}^B$ is the first $\bar{d}$ columns of
$S^B$. As a result, the rewiring becomes approximate, $M \approx
\bar{S}^A (\bar{S}^B)^T$.  But since the first $\bar{d}$ columns
correspond to the largest singular values, the approximation is
relatively accurate even for small $D$ and rapidly converges as $D$
is increased~\cite{LN,HinczBerkerTRG}. With this cutoff, the maximum
size of the tensors is bounded as the TRG procedure is iterated and
we can extract numerically thermodynamic information from flows
within a finite-dimensional space of real-valued tensor elements.

\section{TRG for Quenched Randomness: the bond-diluted Ising model}\label{Ising}

\begin{table*}[ht]
{\small \hfill{}
\begin{tabular}{c |c c c |c c c}
\hline
Spin state & \multicolumn{3}{c|}{Type A} & \multicolumn{3}{c}{Type B}\\
\hline
($s_1$, $s_2$, $s_3$) \tp \bt&$(i_1,i_2,i_3)$ &${T}^A_{i_1 i_2 i_3}$&$D^A_{i_1 i_2 i_3}$& $(i_1,i_2,i_3)$ &${T}^B_{i_1 i_2 i_3}$ & $D^B_{i_1 i_2 i_3}$\\
\hline
$\uparrow \uparrow \uparrow$\tp\bt&111 & $e^{\frac{1}{2}(J_1+J_2+J_3+2H_1+2H_2+2H_3)}$ & $e^{\frac{1}{2}(J_1+J_2+J_3)}$& 111 & $e^{\frac{1}{2}(J_1+J_2+J_3+2H_1+2H_2+2H_3)}$ &$e^{\frac{1}{2}(J_1+J_2+J_3)}$\\
$\uparrow \uparrow \downarrow$\tp\bt&214&$e^{\frac{1}{2}(-J_1+J_2-J_3+2 H_2)}$&$0$&124&$e^{\frac{1}{2}(J_1-J_2-J_3+2 H_1)}$&$e^{\frac{1}{2}(J_1-J_2-J_3)}$\\
$\uparrow \downarrow \uparrow$\tp\bt&142&$e^{\frac{1}{2}(J_1-J_2-J_3+2 H_1)}$&~~$e^{\frac{1}{2}(J_1-J_2-J_3)}$&241&~$e^{\frac{1}{2}(-J_1-J_2+J_3+2 H_3)}$&$0$\\
$\uparrow \downarrow \downarrow$\tp\bt&243&~~$e^{\frac{1}{2}(-J_1-J_2+J_3-2 H_3)}$&$0$&234&~~~$e^{\frac{1}{2}(-J_1+J_2-J_3-2 H_2)}$&$0$\\
$\downarrow \uparrow \uparrow$\tp\bt&421&$e^{\frac{1}{2}(-J_1-J_2+J_3+2 H_3)}$&$0$&412&~$e^{\frac{1}{2}(-J_1+J_2-J_3+2 H_2)}$&$0$\\
$\downarrow \uparrow \downarrow$\tp\bt&324&~$e^{\frac{1}{2}(J_1-J_2-J_3-2 H_1)}$&$-e^{\frac{1}{2}(J_1-J_2-J_3)}$&423&$e^{\frac{1}{2}(-J_1-J_2+J_3-2 H_3)}$&$0$\\
$\downarrow \downarrow \uparrow$\tp\bt&432&~$e^{\frac{1}{2}(-J_1+J_2-J_3-2 H_2)}$&$0$&342&$e^{\frac{1}{2}(J_1-J_2-J_3-2 H_1)}$&$-e^{\frac{1}{2}(J_1-J_2-J_3)}$\\
$\downarrow \downarrow \downarrow$\tp\bt&333&$e^{\frac{1}{2}(J_1+J_2+J_3-2H_1-2H_2-2H_3)}$&$-e^{\frac{1}{2}(J_1+J_2+J_3)}$&333&~~$e^{\frac{1}{2}(J_1+J_2+J_3-2H_1-2H_2-2H_3)}$&$-e^{\frac{1}{2}(J_1+J_2+J_3)}$\\
\hline
\end{tabular}}
\hfill{}
\caption{The tensor elements for the bond-diluted Ising model, as
  defined in Secs.~\ref{mapping} and \ref{spinspin}, for the first renormalization step.  The first column
  gives the spin state $(s_1,s_2,s_3)$ for a triangle of the
  original triangular lattice, following the convention of
  Fig.~\ref{dual}.  For the type A triangle, the next three columns
  show the associated composite indices $(i_1,i_2,i_3)$, and the
  tensor elements $T^A_{i_1i_2i_3}$, $D^A_{i_1i_2i_3}$.  The last
  three columns show the analogous information for the type B
  triangle.  All tensor elements not shown are zero.}
\label{tab1}
\end{table*}

\subsection{The Bond-Diluted Ising Hamiltonian and Its Mapping onto a Tensor Network}\label{mapping}

The general Hamiltonian for a quenched random Ising system is
\begin{equation}\label{i1}
 -\beta { \cal H}  = \sum_{\langle ij \rangle} \left[J_{ij} s_i s_j +  H_{ij} (s_i+s_j)\right] \,, \quad s_i=\pm 1\,,
\end{equation}
\noindent where $\beta=1/k_BT$, $J_{ij}$ and $H_{ij}$ are
respectively the local spin-spin coupling and magnetic field for
sites $i$ and $j$, and $\langle i j \rangle$ denotes a sum over
nearest-neighbor pairs of sites. Although this Hamiltonian
encompasses a variety of models, all the way to the random-field
spin glass~\cite{Migliorini}, we shall here focus on a the
bond-diluted Ising case, where the interaction constants $J_{ij}$
are distributed with a quenched probability ${\cal P}(J_{ij})$ of
the form
\begin{equation}\label{i2}
{\cal P} (J_{ij}) = p \delta (J_{ij}) + (1-p) \delta (J_{ij}-J).
\end{equation}
Here $J > 0$, implying ferromagnetism, and $p$ is the fraction of
missing bonds. While we restrict our attention to the zero magnetic
field subspace, $H_{ij} =0$, formally the local fields will be kept
in the Hamiltonian in order to take derivatives to obtain
thermodynamic functions.

\begin{figure}
\includegraphics*[width=0.5\columnwidth]{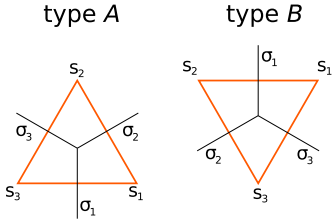}
\caption{(Color online) Duality mapping between spin states on the
triangular lattice and bond variables in the tensor network. The
variables $s_i = \pm 1$ at the triangle corners correspond to Ising
spins in the Hamiltonian of Eq.~\eqref{i1}.  The bond variables
$\sigma_i$ are products of the $s_i$ connected by the bond.  Up and
down triangles yield type A and B tensors respectively.}\label{dual}
\end{figure}

Starting with the Hamiltonian of Eq.~\eqref{i1} on a triangular
lattice, a duality transformation allows us to express the partition
function as a hexagonal-lattice tensor network. (The duality for Potts spins would generate three-point interactions, which would be included in the definition of the tensor $T_{i_1i_2i_3}$.)  Each
triangle in the triangular lattice corresponds to a tensor, with up
triangles associated with a $T^A$ and down triangles with a $T^B$,
as shown in Fig.~\ref{dual}. For spin variables $s_i$, $s_j$, $s_k$
in a given triangle in the manner illustrated in the figure, we
define corresponding edge variables $\sigma_m$ as the products of
neighboring $s_m$, i.e., for the type $A$ triangle, $\sigma_1=s_3
s_1$, $\sigma_2 = s_1 s_2$, $\sigma_3=s_2 s_3$ and for the type B
triangle, $\sigma_1 = s_1 s_2$, $\sigma_2 = s_2 s_3$, $\sigma_3 =
s_3 s_1$.  Since $s_m = \pm 1$ and $\sigma_m = \pm 1$, we can now
introduce a composite index $i_m \equiv (5-\sigma_m-2s_m)/2$ which
runs from 1 to 4 and describes the four possible states of the $m$th
triangle edge. Letting $J_m$ be the coupling $J_{ij}$ associated
with this edge and $H_m = H_{ij}$ be the edge magnetic field, then
the tensors for the two triangles types are:
\begin{equation}\label{i3}
\begin{split}
 T^A_{i_1i_2i_3}=&e^{\frac{1}{2}\left(\sum_{m=1}^3 J_m\sigma_m+H_m(1+\sigma_m)s_m\right)}P(\sigma_1\sigma_2\sigma_3)\\
&\cdot P(\sigma_1s_1s_3)P(\sigma_2s_2s_1)P(\sigma_3s_3s_2)\,, \\
 T^B_{i_1i_2i_3}=&e^{\frac{1}{2}\left(\sum_{m=1}^3 J_m\sigma_m+H_m(1+\sigma_m)s_m\right)}P(\sigma_1\sigma_2\sigma_3)\\
&\cdot P(\sigma_1s_1s_2)P(\sigma_2s_2s_3)P(\sigma_3s_3s_1)\,,
\end{split}
\end{equation}
where $P(x)\equiv (1+x)/2$ is a projection operator.  The $P$
factors in the tensors remove the bond states that do not correspond
to a physically allowable spin configuration.  As a result of the
projection operators, only 8 out of the 64 elements in the tensor
are nonzero. These are listed, for the first renormalization step,
in the 3rd and 6th columns of Table~\ref{tab1} for $T^A$ and $T^B$
respectively.

\subsection{Local Magnetization and Spin-Spin Correlation Function}\label{spinspin}

In order to derive expressions for thermodynamic quantities in the
tensor formalism, let us now restrict the notation $T^A$ and $T^B$
to tensors in the zero magnetic field subspace.  We place a local
magnetic field $H_k$ only at a single location $k$. Let us call the
two tensors which share this bond $\widetilde T^A$ and $\widetilde
T^B$.  These are the only two tensors in the system whose components
are modified by the local field.  The corresponding partition
function is
\begin{equation}\label{i4}
Z = \sum_{i_1,\ldots,i_K} T^A_{i_1 i_2 i_3}T^B_{i_4 i_5 i_3} \cdots
\widetilde T^A_{i_k i_l i_m}\widetilde T^B_{i_k i_n i_o} \cdots
T^B_{i_{K-2} i_{K-1} i_K}.
\end{equation}
Without loss of generality we take the contraction of the
$\widetilde T^A$ and $\widetilde T^B$ tensors to be Case 2 in
Eq.~\eqref{t1}, since the derivation proceeds analogously for the
other Cases.

The local magnetization is $m_k = \langle (s_i + s_j )/2 \rangle
\equiv \langle S_k \rangle$ for the sites $i$, $j$ associated with
the bond $k$.  In terms of the local magnetic field $H_k$, the
magnetization $m_k$ is given by the derivative
\begin{equation}\label{i5}
\begin{split}
m_k =&\left.\frac{1}{2}\frac{\partial\ln Z}{\partial
H_k}\right|_{H_k=0}\\
=& \frac{1}{2 Z} \sum_{i_1,\ldots,i_K} \left\{ T^A_{i_1 i_2 i_3}T^B_{i_4 i_5 i_3} \cdots D^A_{i_k i_l i_m} T^B_{i_k i_n i_o} \cdots \right.\\
&+ \left.T^A_{i_1 i_2 i_3}T^B_{i_4 i_5 i_3} \cdots T^A_{i_k i_l i_m}
D^B_{i_k i_n i_o} \cdots \right\}\,,
\end{split}
\end{equation}
where the differentiated tensors are
\begin{equation}\label{i6}
\begin{split}
D^A_{i_{k} i_{l} i_{m}} =& \left. \frac{\partial\widetilde{T}^A_{i_{k} i_{l} i_{m}}}{\partial H_{i_{k}}} \right|_{H_{i_{k}}=0}, \quad D^B_{i_k i_n i_o} = \left. \frac{\partial\widetilde{T}^B_{i_{k} i_{n} i_{o}}}{\partial H_{i_{k}}} \right|_{H_{i_{k}}=0}. \\
\end{split}
\end{equation}
The nonzero elements of $D^A$ and $D^B$ are shown, for the first
renormalization step, in the 4th and 7th columns of
Table~\ref{tab1}.

After taking the average over the disorder, the first and second
terms in the brackets on the right-hand side of Eq.~\eqref{i5} are
equivalent, so that
\begin{equation}\label{i7}
\begin {split}
[ m_k] =& [\langle S_k \rangle]=\\
& \biggl[Z^{-1} \sum_{i_1,\ldots,i_K} T^A_{i_1 i_2 i_3}T^B_{i_4 i_5
i_3} \cdots D^A_{i_k i_l i_m} T^B_{i_k i_n i_o} \cdots \biggr].
\end{split}
\end{equation}
A similar derivation for the correlation function yields
\begin{equation}\label{i8}
\begin{split}
[\langle S_k S_l \rangle] =& \biggl[Z^{-1} \sum_{i_1,\ldots,i_K} T^A_{i_1 i_2 i_3}T^B_{i_4 i_5 i_3} \cdots D^A_{i_k i_l i_m} T^B_{i_k i_n i_o} \cdots \\
& \cdots D^A_{i_l i_p i_q}T^B_{i_l i_r i_s} \cdots \biggr].
\end{split}
\end{equation}
We shall be interested in long-range correlations, as an indicator
of thermodynamic phase behavior.  In this case, the four individual
$s_i$ spin-spin correlations that make up the $[\langle S_k S_l
\rangle]$ are approximately equal: $[\langle S_k S_l \rangle]
\approx [\langle s_i s_j \rangle]$, where $s_i$ is either of the
spins contributing to $S_k$ and $s_j$ is either of the spins
contributing to $S_l$.  Hence we shall use $[\langle S_k S_l
\rangle]$ and $[\langle s_i s_j \rangle]$ interchangeably in the
rest of the text.

\subsection{Details of the Numerical Implementation}

To calculate the long-range spin-spin correlation function $[\langle
  S_k S_l \rangle]$, we start with a finite hexagonal lattice after
$n$ construction steps, with size varying between $n=7-10$ steps
($N=17496 - 472392$ tensors).  The bonds $k$ and $l$ are chosen to
be at the maximum separation within the lattice, taking periodic
boundary conditions into account.  For a given realization of the
disorder, the sum on the right-hand side of Eq.~\eqref{i8} is
evaluated by doing $n$ TRG transformations, which yields the
contraction in terms of four renormalized tensors in the $n=0$
structure.  These last four tensors are directly contracted.  A
similar process yields the value of the partition function $Z$ which
is the denominator in Eq.~\eqref{i8}.  The configurational average
is taken over $200-300$ realizations, implemented by randomly
assigning the $J_{ij}$ on the initial lattice according to the
probability distribution in Eq.~\eqref{i2}. The tensors on the
original lattice, i.e., in Eqs.~\eqref{i3} and \eqref{i6}, have
index range $d=4$.  For subsequent tensors, we use a cutoff
parameter $D=8-14$.

Some tensor elements tend to grow exponentially in magnitude as the
TRG transformation is iterated, which poses potential numerical
difficulties.  To counteract this, we take advantage of the fact that
we can always factor out a constant from each tensor without changing
the physics.  For each tensor during each TRG iteration, the factor
extracted is equal to $\text{min}(T_\text{max},2)$ where $T_\text{max}$ is
the maximum absolute value of the tensor elements.  Keeping an upper
bound of 2 on this extracted factor slows down the decay of most
tensor elements to zero, which would otherwise lead to other numerical
artifacts.  We keep track of the total extracted factors in the
numerator and denominator of Eq.~\eqref{i8}, which are then used in
calculating the final correlation function value.

\begin{figure}
\includegraphics*[width=\columnwidth]{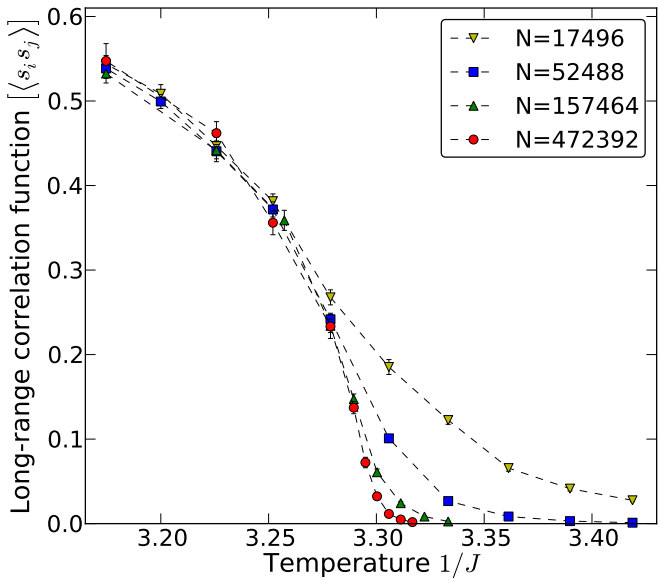}
\caption{\label{fig3}(Color online) The long-distance spin-spin
correlation $[\langle s_i s_j \rangle]$ as a function of temperature
$1/J$, calculated using the TRG method for bond dilution probability
$p=0.1$ and cutoff parameter $D=8$.  The curves for four different
initial tensor network sizes $N$ are shown.}
\end{figure}

\begin{figure}
\includegraphics*[width=\columnwidth]{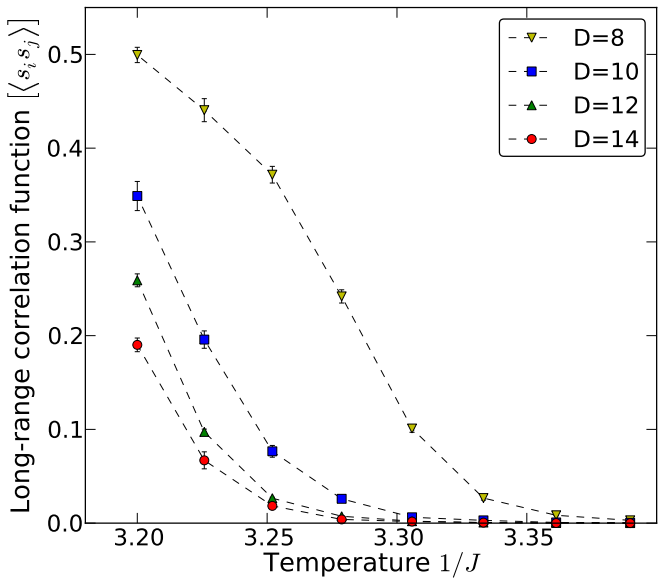}
\caption{\label{fig4} (Color online) The long-distance spin-spin
correlation $[\langle s_i s_j \rangle]$ as a function of temperature
$1/J$, calculated using the TRG method for bond dilution probability
$p=0.1$ and network size $N=157464$ tensors.  The curves for four
different cutoff parameters $D$ are shown.}
\end{figure}

\section{Results}\label{results}

Representative results for the long-distance spin-spin correlation
function $[\langle s_i s_j \rangle]$ as a function of temperature
$1/J$ at bond dilution $p=0.1$ are given in Figs.~\ref{fig3} and
\ref{fig4}.  The former shows curves for various tensor network
sizes $N$ using cutoff $D=8$, while the latter varies the cutoff $D$
at fixed size $N=157464$.  Away from the critical temperature, where
widely separated spins are uncorrelated, $[\langle s_i s_j \rangle]
\approx [\langle s_i \rangle^2]$, and we expect distinct limiting
behaviors for the two different thermodynamic phases in the system:
at low $1/J$ in the ferromagnetically ordered phase $[\langle s_i
s_j \rangle] \to 1$, while at high $1/J$ in the paramagnetic phase
$[\langle s_i s_j \rangle] \to 0$.  The temperature region where one
sees a smooth transition between these two regimes for finite
systems, illustrated in Figs.~\ref{fig3} and \ref{fig4}, gives a
rough indication of the phase transition temperature $1/J_c$ in the
thermodynamic limit.  With increasing $N$ in Fig.~\ref{fig3} and
increasing $D$ in Fig.~\ref{fig4}, the transition becomes sharper,
as our truncations converge toward the exact result for an infinite
system. The probability $p=0.1$ at which these results are
calculated is smaller than the threshold value $p_c \approx
0.653$~\cite{Stauffer}, above which the triangular lattice no longer
percolates.  For $p > p_c$ we would not see a transition region: the
paramagnetic phase exists at all temperatures, since islands of
ordered spins of size $\sim {\cal O}(N)$ become exponentially
improbable.

\begin{figure}[t]
\includegraphics*[width=\columnwidth]{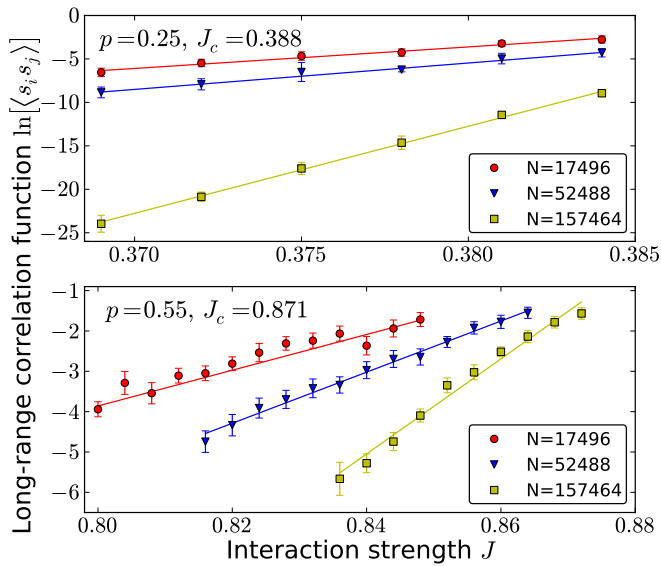}
\caption{\label{fig5} (Color online) Data points show the log of the
long-distance spin-spin correlation, $\ln [\langle s_i s_j
\rangle]$, as a function of interaction strength $J$ for three
different system sizes $N$ and two different bond dilution
probabilties $p$ (top panel: $p=0.25$, bottom panel: $p=0.55$).  The
weighted least squares linear fits, shown as solid lines, yield the
coefficients $A_i$ and $B_i$ in Eq.~\eqref{r2}, which allow one to
estimate $J_c$ through finite size scaling, Eq.~\eqref{r3}.  The
resulting values of $J_c$ are 0.388 ($p=0.25$) and 0.871
($p=0.55$).}
\end{figure}

To obtain an accurate estimate of the exact transition temperature
$1/J_c$, we can employ the following finite-size scaling relation,
which describes the ratios of the correlation functions at three
different system sizes $N_1$, $N_2$, and $N_3$ when
$J=J_c$~\cite{Takano}:
\begin{equation}\label{r1}
\frac{\ln
\left(\frac{g(N_2)}{g(N_1)}\right)}{\ln\left(\frac{N_2}{N_1}\right)}=\frac{\ln\left(\frac{g(N_3)}{g(N_2)}\right)}{\ln\left(\frac{N_3}{N_2}\right)}\,,
\end{equation}
where $g(N)$ is the long-distance correlation function $[\langle
  s_is_j\rangle]$ for network size $N$.  For the $i$th system, at the temperature region where $g(N_i)$ decays rapidly to zero ($J$
just smaller than $J_c$), the decay is approximately exponential in
$J$,
\begin{equation}\label{r2}
\ln(g(N_i)) \approx A_i J - B_i,
\end{equation}
for some constants $A_i$ and $B_i$.  This exponential behavior for
three different system sizes is shown in Fig.~\ref{fig5} for
$p=0.25$ and $0.55$.  To calculate $A_i$ and $B_i$, we do a weighted
linear least squares fit to $\ln(g(N_i))$ vs. $J$ data in a region
of $J$ where the relative uncertainty (from the configurational
average) for the data points is less than $15\%$. Plugging
Eq.~\eqref{r2} into Eq.~\eqref{r1} with $J=J_c$, we can solve for
$J_c$ in terms of the $A_i$, $B_i$, and $N_i$,
\begin{equation}\label{r3}
J_c= \frac{(B_{2}-B_{1}) \ln
   \left(\frac{N_{3}}{N_{2}}\right)+(B_{2}-B_{3}
   ) \ln
   \left(\frac{N_{2}}{N_{1}}\right)}{(A_{1}-A_{2
   }) \ln
   \left(\frac{N_{3}}{N_{2}}\right)+(A_{3}-A_{2}
   ) \ln \left(\frac{N_{2}}{N_{1}}\right)}.
\end{equation}

Carrying out this calculation across the entire $p$ range for
$N_1=17496$, $N_2=52488$, and $N_3=157464$ at $D=12$, we obtain the
phase diagram shown in Fig.~\ref{fig6}.  For comparison we also plot
the same phase diagram obtained from a rigorous approximation scheme
for the bond-diluted Ising model free energy~\cite{Georges}, which
can be considered exact on the scale of the figure.  The agreement
is quite close, with an average relative deviation of 1\%.  Two
values along the curve are known exactly: $1/J_c = 4/\ln 3 =
3.641$~\cite{Wannier} at $p=0$ and the curve goes to $1/J_c=0$ at
the percolation treshold $p = p_c = 1 - 2 \sin(\pi/18) =
0.653$~\cite{Stauffer}.  Our results deviate from these exact values
by 0.3\% and 0.4\% respectively.

\section{Conclusions}\label{conclusions}

We have shown how the TRG approach provides an efficient and precise
method for calculating thermodynamic properties of a quenched random
classical model---the triangular-lattice bond-diluted Ising
Hamiltonian.  By expressing the partition function and related
quantities such as spin-spin correlation functions in terms of
tensor networks, they can be readily evaluated through TRG for large
lattice sizes.  In combination with finite-size scaling ideas, the
result is a precise estimate of the phase diagram.  If desired,
convergence to the exact critical properties can be achieved by
increasing the cutoff parameter defining the index range of the
tensors.

The bond-diluted Ising model is only a first step in the exploration
of disordered systems using TRG: the methods presented here are
easily extended to frustrated Hamiltonians exhibiting spin-glass
behavior and the resulting complex multicritical phase structures.
The numerical accuracy of the technique will be a valuable feature
in probing analytical conjectures on the exact locations of
spin-glass multicritical
points~\cite{Nishimori,HinczBerkerMC,Ohzeki1,Ohzeki2}.

\begin{acknowledgments}
This research was supported by the Alexander von Humboldt
Foundation, the Scientific and Technological Research Council of
Turkey (T\"UBITAK), and the Academy of Sciences of Turkey.
Computational resources were provided by the Gilgamesh cluster of
the Feza G\"ursey Research Institute.
\end{acknowledgments}

\end{document}